\documentclass[onecolumn,preprintnumbers,superscriptaddress,amsmath,amssymb,aps]{revtex4}
\usepackage{graphicx,float}
\usepackage{comment}

\begin{document}

\title{Influence of an environment changing in time on Crucial Events : the earthquake prototype}

\author{Callum Muir}
\affiliation{Center for Nonlinear Science, University of North Texas, P.O. Box 311427, Denton, Texas 76203-1427}
\author{Mauro Bologna}
\affiliation{Departamento de Ingeniería Eléctrica-Electrónica, Universidad de Tarapacá, Arica, Chile}
\author{Paolo Grigolini}
\affiliation{Center for Nonlinear Science, University of North Texas, P.O. Box 311427, Denton, Texas 76203-1427}

\begin{abstract}
This paper is devoted to the study of the interaction between two distinct forms of non-stationary processes, which we will refer to as non-stationarity of first and second kind.  The non-stationarity of first kind is caused by criticality-generated events that we call \emph{crucial events}. Crucial events signal ergodicity breaking emerging from the interaction between the units of the complex system under study, indicating that the non stationarity of first kind has internal origin. The non-stationarity of second kind is due to the influence on the system of interest of an environment changing in time, thereby implying an external origin. In this paper we show that the non-stationarity of first kind, measured by an inverse power law index $\mu$ is characterized by singularities at $\mu =2$ and $\mu = 3$. We realize the interaction between the non-stationarity of first kind and the non-stationarity of second kind with a model frequently adopted to study earthquakes, namely, a system of main-shocks, assumed to be crucial events, generating a cascade of after-shocks simulating the changing in time environment. We prove that the after-shocks significantly affects the detection of anomalous scaling, with this effect weakening as the value $\mu$ approaches $\mu = 2.5$. We argue that this  result is a consequence of the fact that the states $\mu = 2$ and $\mu = 3$  are the borders between different statistical regimes, where a sort of phase transition occurs, with $\mu =2.5$ being a state  sufficiently far from both transition regimes. We conclude this paper with the observation that the earthquakes should be interpreted as resulting from the interaction between many geophysical units generating criticality, with the non-stationary events of second kind affecting conveniently short time regions between two consecutive crucial events.

 \end{abstract}

\maketitle

\section{Introduction} \label{Introduction}

The earthquake issue in the recent past has been the subject of an intense debate concerning the power-law behavior in time distribution of earthquakes  \cite{doublesornette},
\cite{helmstetter},  \cite{bak}, \cite {mega}, \cite{turcotte}, \cite{kagan}.  This problem goes much beyond geophysics insofar as it is expected to be a property of complex systems, including bacterial  persistence \cite{bacterial}, solar radiophysics \cite{solar}, stock price
fluctuations \cite{stok}, tree-limb branching \cite{tree} and the time duration of dictatorships \cite{romanempire}. The main source of this debate is the fact that the earthquakes are supposed to be generated by main shocks activating a cascade of aftershocks. The authors of Ref. \cite{mega} made the conjecture that the main shocks are crucial events. This assumption implies that they are not stationary. 

The concept of deviation from stationarity is an issue of increasing interest in the field of complexity. We focus on experimental data expressed under the form of a time series, $\{\xi(t)\}$. The autocorrelation function 

\begin{equation}
\Phi_{\xi}(t_1,t_2)  = <\xi(t_1)\xi(t_2)>
\end{equation}
in general depends on both $t_1$ and $t_2$. The symbol $<...>$ represents a Gibbs average of the quantity indicated by the three dots, $\xi(t_1)\xi(t_2)$ in this case. 
The stationarity condition is the idealized mathematical prescription:
\begin{equation}
\Phi_{\xi}(t_1,t_2)  =  \Phi(|t_1 - t_2|) , 
\end{equation}
with the symbol $\Phi(t)$ denoting a stationary correlation function. This stationarity condition 
 can be violated leading to an extended time regime where the correlation function requires an explicit dependence on both $t_1$ and $t_2$.  The deviation from the stationarity condition can be due to two distinct reasons: (1) {\bf Non-stationarity of first kind.} The complex system under study is characterized by aging, namely, it has a dynamics strongly dependent on the initial condition and on the time distance from the initial condition. The dependence on the initial condition vanishes very slowly as time increases; (2)  {\bf Non-stationarity of second kind.} The system might move towards the stationary condition, but the external environment
influencing the dynamics of the system of interest  changes in time.  

It is interesting to notice that the joint action of these forms of non-stationarity was recently discussed by Kataoka et al \cite{akimoto}, although reversing our definition, these authors denote the non-stationarity of first kind as non-stationarity of second kind and the non-stationarity of second kind as a non-stationarity of first kind.  

We note that the non-stationarity of first kind is closely connected to the engineering view of Cox \cite{cox}, who defined the age dependent failure rate
\begin{equation} \label{cox}
g(t) = \frac{\psi(t)}{\Psi(t)}, 
\end{equation}
where $\psi(t)$ is the waiting time distribution density between the birth of a machine and the occurrence of its first failure, and $\Psi(t)$ is the probability that no failure occurs up to time $t$. 

Since 
\begin{equation}
\psi(t) =  -\frac{d}{dt} \Psi(t),
\end{equation}
the use of Eq. (\ref{cox}) 
yields

\begin{equation}
\Psi(t) = exp{\left\{-\int_{0}^t dt' g(t')\right\}}. 
\end{equation}

It is straightforward to prove that the assumption

\begin{equation}
g(t) = \frac{r_0}{1 + r_1 t}
\end{equation}

yields

\begin{equation}  \label{manneville}
\Psi(t) = \left(\frac{T}{t+T}\right)^{\mu-1}, 
\end{equation}
where
\begin{equation}
\mu = 1 + \frac{r_0}{r_1}, 
\end{equation}
with the corresponding survival probability 

\begin{equation}  \label{ordinary}
\psi(t) = (\mu-1) \frac{T^{\mu-1}}{(t+ T)^{\mu}}. 
\end{equation} 

It is important to notice that if we generate a large number of sequences with the first event occurring at time $t= 0$ the rate of events at a generic time $t > 0$, called $R(t)$ 
is \cite{feller} 
\begin{equation}  \label{perennialaging}
R(t)  = \frac{1}{t^{2-\mu}}, 
\end{equation}
if $1 < \mu < 2$ and \cite{bellazzini}
\begin{equation} \label{formauro}
R(t) =  \frac{1}{<\tau>} \left [1 +  \frac{1}{3-\mu}\left(\frac{T}{t}\right)^{\mu-2}\right]
\end{equation}
with
\begin{equation} \label{formauro2}
<\tau> = \frac{T}{\mu-2},
\end{equation}
if $2 < \mu < 3$. 

The condition $\mu = 2$ corresponds to a transition from the region where the rate of crucial events is never constant to the region where the rate of crucial events becomes constant in the long time limit.
 This transition has been widely discussed in literature. See, for example, \cite{barkai1,barkai2,barkai3,barkai4,barkai5}.  The transition from $\mu < 3$ to $\mu >3$
corresponds to the transition from the regime characterized by L\'{e}vy statistics to the regime characterized by Poisson statistics. It is an important transition reminiscent of phase transition processes \cite{annunziato}.

The main goal of this manuscript is to study the interaction between non-stationarity of first kind and non-stationarity of second kind. 

The outline of this paper is as follows. We devote Section \ref{review} to discuss the physical meaning of $\mu =2$ and $\mu =3$. In Section \ref{omori} we illustrate a model of aftershock that we use in this paper to realize the non-stationarity of second kind and in Section \ref{interaction} we make a statistical analysis of the interaction between the two forms on non-stationarity. We devote Section \ref{last} to concluding remarks emphasizing that the results of this paper should be an incentive to explore the still poorly understood terrestrial underground dynamics.

\section{Review of the special properties of the condition $\mu = 2$ and $\mu=3$} \label{review} 

\subsection{From $\mu <2$ to $\mu> 2$}

There exist two main proposals to describe the waiting time distribution density with $\mu < 2$. The first is 
the waiting time distribution density  $\psi(t)$ corresponding to the survival probability of Eq. (\ref{manneville}).
The second is the Mittag-Leffler survival probability $\Psi_{ML}(t)$ defined by its Laplace transform

\begin{equation}
\hat{\Psi}_{ML}(s) = \frac{1}{s + \lambda^{\alpha} s^{1-\alpha}}, 
\end{equation}
with $0 < \alpha <2$

The corresponding waiting time distribution density reads
\begin{equation}  \label{indistinguishable}
\hat{\psi}_{ML}(s) = \frac{1}{1 +  \left (\frac{s}{\lambda }\right)^{\alpha}}. 
\end{equation}

Note that the Laplace transform of $\psi(t)$ of Eq. (\ref{ordinary}) for $s\rightarrow 0$ is:
\begin{equation}
\hat{\psi}(s) = 1 - \Gamma(1-\alpha)(Ts)^{\alpha} +.....,
\end{equation} 

which for $s\rightarrow 0$ is indistinguishable from Eq. (\ref{indistinguishable}) if

\begin{equation}
T = \frac{1}{\lambda} \frac{1}{\Gamma(1-\alpha)^{1/2}}. 
\end{equation}

Moving to $\alpha > 1$ the ML becomes a correlation function with values non-necessarily positive, while Manneville keeps the survival probability condition.

If we limit our attention to $\Psi(t)$ of Eq. (\ref{manneville}) and to corresponding waiting time distribution density  of Eq. (\ref{ordinary}) we are apparently led to believe that this is a transition 
from  perennial aging of Eq. (\ref{perennialaging}) to the condition of Eq. (\ref{formauro}), where non-stationary of first kind is temporary. 
However, the main focus of this paper is the region $2 < \mu <3$, 
where the correlation function becomes stationary in the long-time limit. However, despite the recovery of stationarity in the long-time limit, this regime is characterized by deviations from ordinary statistical physics, including multifractality \cite{tenyearsearlier}, ergodicty breaking and infinite density \cite{hanngy, metzler}. 

In this paper we find an additional anomalous property of the region $2 < \mu < 3$. In fact, the property $R(t)$ defined by Eq. (\ref{formauro})  and Eq. (\ref{formauro2}) for $\mu$ tending to the singularity of $\mu = 3$ is characterized by an important property. 
Eq.(\ref{formauro}) is an approximated expression of the exact expression
\begin{equation}
R(t) = \sum_{n=1}^{\infty} \psi_n(t),
\end{equation}
where $\psi_n(t) $ is the probability that the $n-th$ crucial event following the first event occurring at time $t = 0$, corresponding to $\psi_0(t) = \delta (t)$, occurs at time $t> 0$. 

It is important to note that crucial events are renewal events, whose occurrence at a given time $t$ does not have any memory of the times of occurrence of the earlier crucial events, with the nature of those crucial events being taken into account by:
\begin{equation}
    \psi_n(t) = \int_{0}^t dt' \psi_1(t') \psi_{n-1}(t-t').
\end{equation}
Using the property that the Laplace transform of a convolution between two time function is the product of the Laplace transform of the two functions, we get 
\begin{equation}
    \hat{R}(s) = \sum_{n=1}^{\infty} \hat \psi_n(s)  = \sum_{n=0}^{\infty} \hat \psi_n(s) -1 = \frac{\hat{\psi}(s)}{1 - \hat{\psi}(s)},
\end{equation}
leading to
\begin{equation}
\hat{R}(s) =  \frac{\hat{\psi}(s)}{1 - \hat{\psi}(s)}.
\end{equation}
Note that $\psi_0(t) = \delta(t) $ and $\psi_1(t) = \psi(t)$.

Using the method illustrated in Appendix A to bypass the singularity of the Gamma function at $\alpha = 1$, we get:
\begin{equation} \label{logarithmicevolution}
R(t) \approx  \frac{1}{ln (\frac{t}{T})}. 
\end{equation}
This implies $R(t)$ does not become constant but undergoes an extremely slow decay.

\subsection{From $\mu < 3$ to $\mu> 3$} \label{3B}

There exists another important effect that has been pointed out in Ref. \cite{annunziato}. Let us assume that the time intervals generated by waiting time distribution density of
Eq. (\ref{ordinary}) are filled with either $W$ or $-W$ by tossing a fair coin.  We generate a time series $\xi(t)$  and we use the time series to create the diffusion process $x(t)$, according to the prescription
\begin{equation} \label{nofriction}
\dot x = \xi(t).
\end{equation}
The Fourier transform of the probability distribution density $p(x,t)$, denoted by the symbol $\hat{p}(k,t)$ is described by the L\'{e}vy prescription
\begin{equation}
\hat{p}(k,t) = e^{-b|k|^{\alpha} t},
\end{equation}
where 
\begin{equation}
\alpha = \mu-1
\end{equation} 
and
\begin{equation}
b =  W\left[(\mu-2)TW)^{\mu-2} sin(\frac{\pi (\mu-2)}{2})\right] \Gamma(3-\mu). 
\end{equation}

We see that if we approach $\mu = 3$ from the side of $\mu < 3$, there a singularity generated by the Gamma function.

Let us now consider the condition $\mu > 3$. In this case the mean waiting time  given by the waiting time distribution density of Eq. (\ref{ordinary}) if finite
and it is given by
\begin{equation}
<t> = \frac{T}{(\mu-2)}. 
\end{equation}
However, also for $\mu > 3$ there are signs of a phase transition generating divergences. This was discussed in Ref. \cite{annunziato} studying the effect of friction through
equation 
\begin{equation}
\dot x = - \lambda x + \xi(t),
\end{equation}
which for $\lambda = 0$ recovers Eq. (\ref{nofriction}).  In this case we get an equilibrium distribution with variance
\begin{equation} \label{variance}
\sigma^2_{\lambda} =  \frac{(\mu-2) W^2 T}{\lambda(\mu-3)}. 
\end{equation}

Notice according to the Appendix, 
\begin{equation} \label{equationofmauro}
R(t) = 1 + \frac{T}{t}.
\end{equation}
This is a very slow transition to the condition of constant rate of events. However, according to Eq. (\ref{formauro}) the condition corresponding to $\mu = 2.5$ generates a regression to constant rate even slower than $1/2$. Therefore, we conclude that the singularity condition generating the result of this paper, concerning the detection of crucial events, perturbed by the Omori cascade is due to the variance singularity of Eq. (\ref{variance}), which makes it difficult to make a distinction between crucial events and non-stationarity of second kind.

\section{A model for earthquake cascade generated by crucial events} \label{omori}
For our model we combine the methods of generating aftershock sequences, assigning an intensity (magnitude) to them, and then spacing both the aftershocks and main shocks by their own unique time distances generated by non-stationary waiting time distributions. For the aftershock generation portion of the model we choose to use the equation from \cite{param}
\begin{equation}\label{modeleq}
\lambda(t,M,M_m) = \frac{10^{a(M_m)-b(M_m)*M}}{(t+c)^{p(M_m)}},
\end{equation}
which is based on the original equation of \cite{gutenomori}.

\begin{figure}
	\begin{center}
		\includegraphics[width=0.4\linewidth]{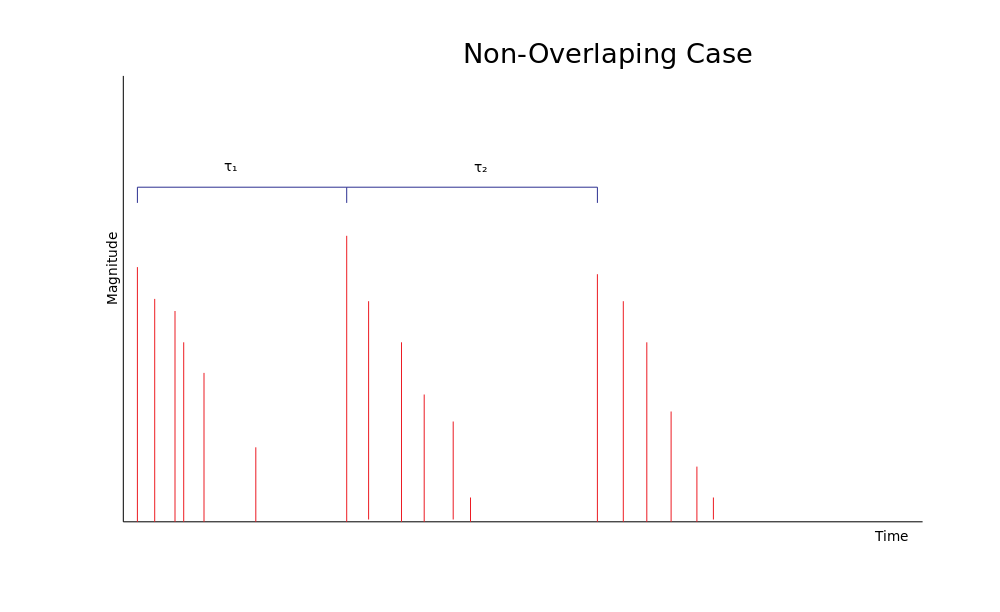}
		\includegraphics[width=0.4\linewidth]{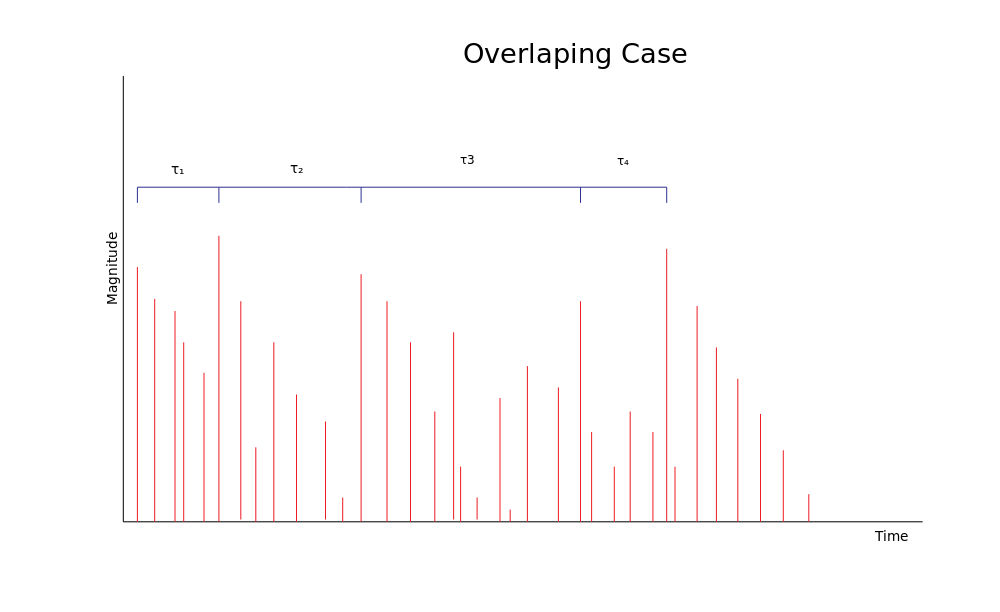}
		\caption{representation of series with overlapping and non-overlapping aftershocks of the model used.}
		\label{overlaps}
	\end{center}
\end{figure}

This equation combines the aftershock rate given by the Omori Law \cite{omori}, \cite{utsu} and the exponential distribution of aftershock intensities given by the Gutenberg-Richter Law \cite{gutenberg}. The parameters $a{(M_m)}$, $b(M_m)$, and $p(M_m)$ are equations that can be fitted from real data. The parameter t represents the time after a main shock occurs, M is an earthquake magnitude lower than that of the initial main shock, and $M_m$ is the intensity of the main shock that caused the aftershock sequence. This portion gives us information on how many aftershocks of intensity M should occur after a main shock and how they should be distributed. For our analysis we limit the magnitude of a main shock $M_m$ to be between the values of 6 and 9. We use the fittings from the authors of \cite{param} in order to get values for the equations in the model. These parameters are given by,
\begin{equation}
	\begin{matrix}
a(M_m) = (0.58+ .12M_o )*M_m - (2.34+ .063*M_o) \\
b(M_m) = 0.12*M_m - 0.063 \\
p(M_m) = -0.06*M_m + 1.44 ,
	\end{matrix}  
\end{equation}
where these linear equations are fitted from data from their analysis of real earthquake data with $M_o$ representing a lower threshold where a shock intensity is no longer considered an aftershock. For our analysis we use the value of 0.1 for $M_o$ and keep the value of c fixed at $10^4$. We describe the time distances between aftershocks with a non-stationary Poisson process, noting that our rate $\lambda$ changes in time due to the environment. After we substitute the fittings into equation (\ref{modeleq}) we can create multiple of these avalanches and space them by a time $\tau$, which is derived from our waiting time distribution density of eq. \ref{ordinary}. We get our times by picking a random number y, which is randomly generated between 0 and 1, and use the equation
\begin{equation} \label{space}
\tau = T*(\frac{1}{y^\frac{1}{\mu-1}}-1)
\end{equation}
to generate our waiting times.
We can create as many of these waiting times as needed in order to space the amount of main shocks we choose to generate. Additionally we can adjust the value of T to space them so that the aftershock sequences can overlap or not (as shown in FIG. \ref{overlaps}), allowing us to create a dense combination of non-stationary processes or allow the processes to remain separated to an extent. For this study we create 100 main shocks with randomly generated magnitudes within our set range to have a sufficient amount for our analysis. Additionally, we vary the value of $\mu$ for both methods of creating our sequences.

\section{Interaction between non-stationarity of the first kind and non-stationarity of the second kind} \label{interaction}
\begin{figure}
	\begin{center}
		\includegraphics[width=0.48\linewidth]{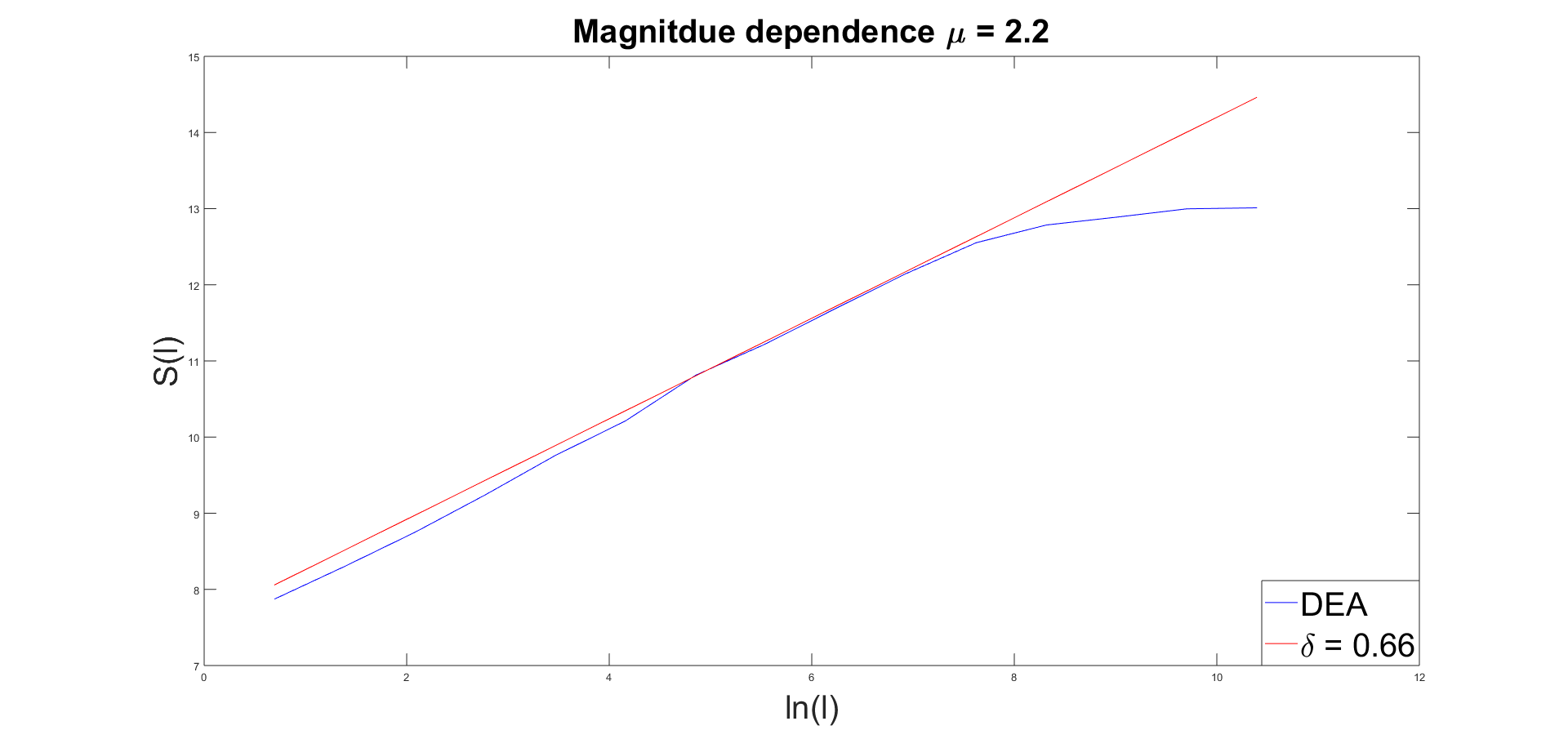}
		\includegraphics[width=0.48\linewidth]{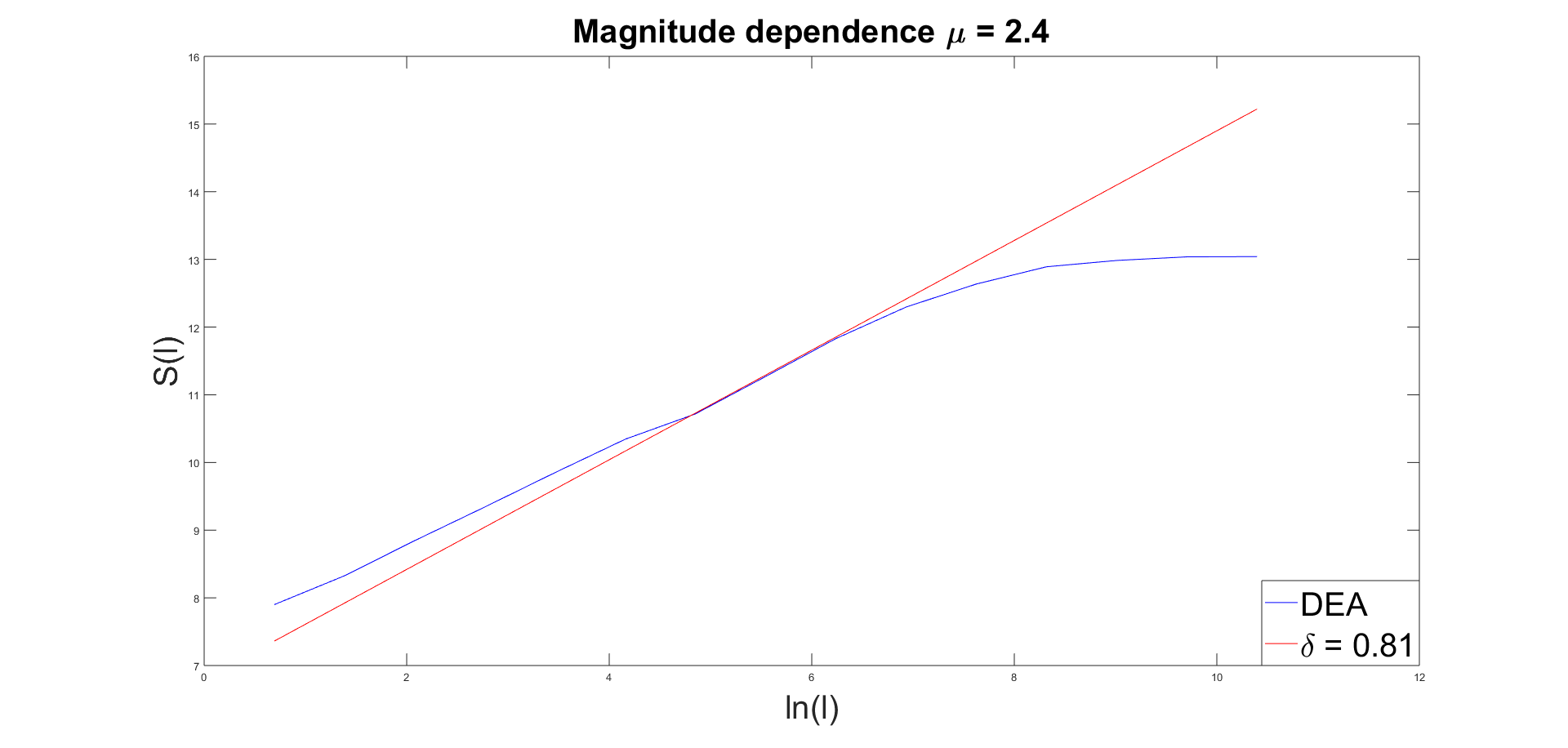}
		\includegraphics[width=0.48\linewidth]{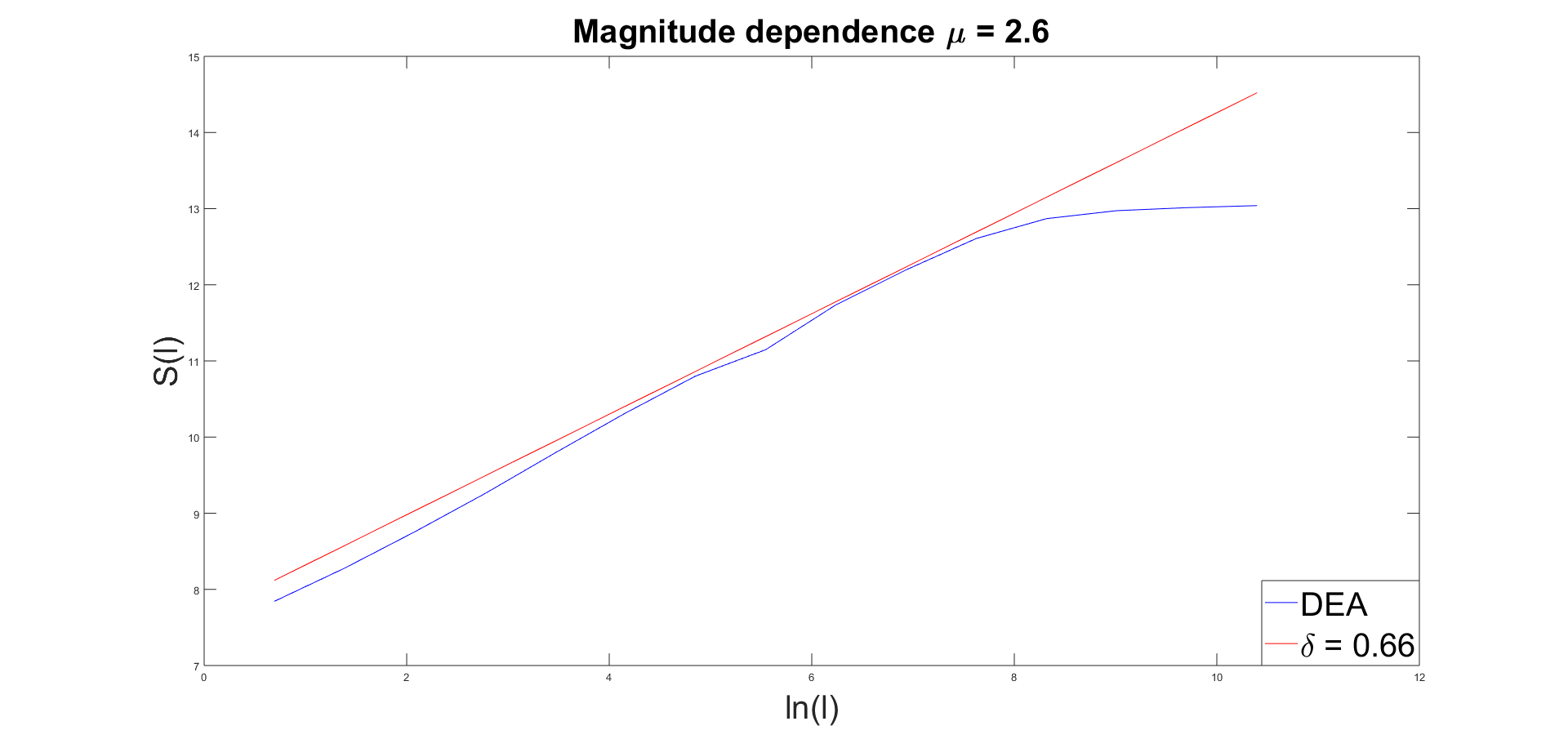}
		\includegraphics[width=0.48\linewidth]{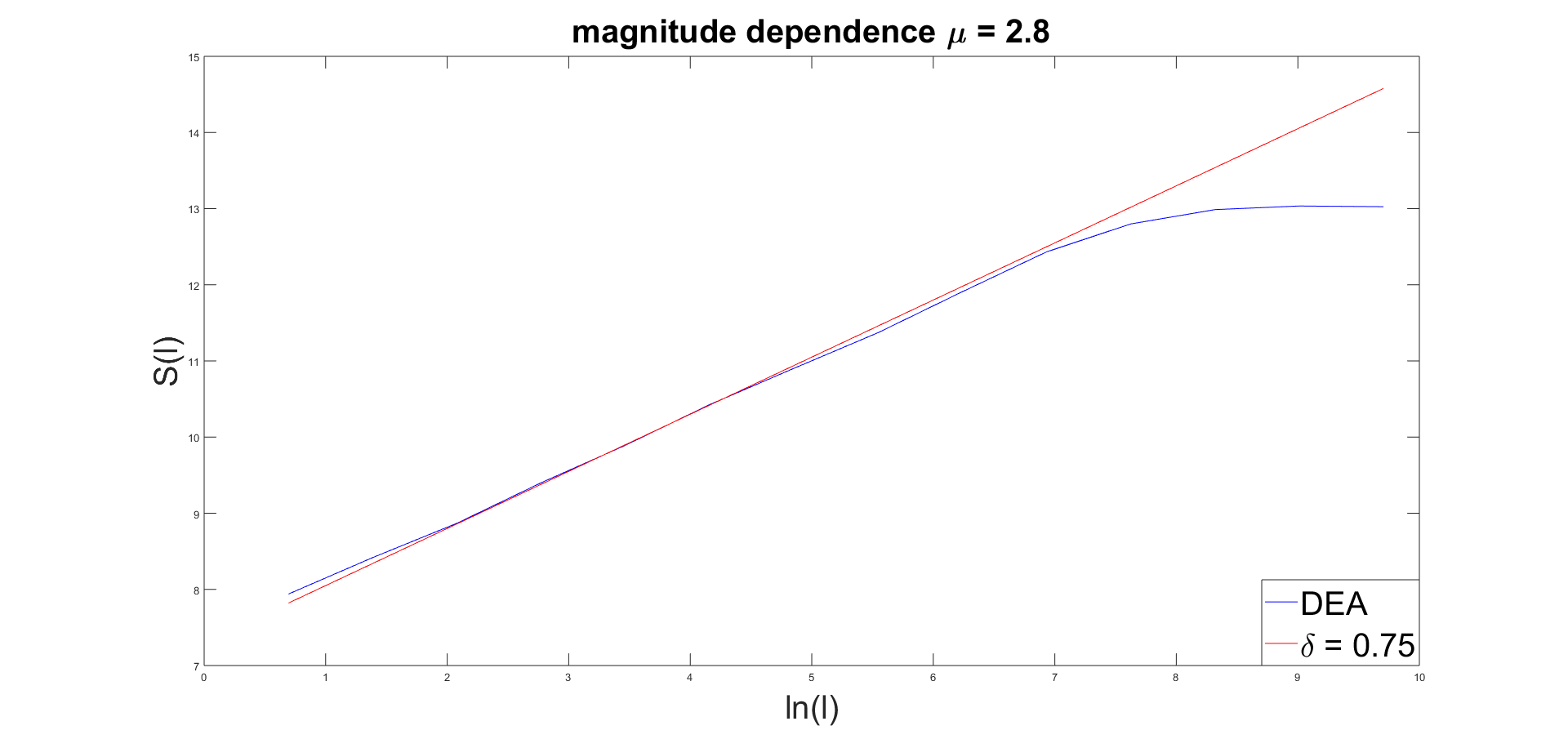}
		\caption{Analysis of various initial values of $\mu$ for overlapping sequences of aftershocks. Titles include the initial value of $\mu$ and the legend provides the resultant scaling in a log-log plot. All of these produce a scaling inconsistent with the initial $\mu$ that should be reproduced through the analysis.}
		\label{overlap}
	\end{center}
\end{figure}

We analyze the series constructed by our model in both the case of overlapping and non-overlapping Omori cascades. Both cases have the effect of hiding the occurrence of main shocks, representing non-stationarity of first kind. However, with non-overlapping cascades we find that the detection of events can be recovered for regions away from the borders of statistical phase changes.
To discover the "invisible" crucial events we adopt the technique of Diffusion Entropy Analysis (DEA) with stripes \cite{dea}. Since this method was created to detect potentially invisible crucial events from real data sets that may contain many other non-crucial events. This analysis takes advantage of creating many diffusion trajectories through use of moving a mobile window of size l and forming the probability density function
\begin{equation}
p(x,l) = \frac{1}{l^\delta}F(\frac{x}{l^\delta})
\end{equation}
from data and evaluating the Shannon Entropy given by
\begin{equation}
S(l) = \int_{-\infty}^\infty p(x,l)\ln{p(x,l)}dx.
\end{equation}
Integration of this equation yields
\begin{equation}
S(l) = A + \delta\ln{(l)},
\end{equation}
where A is a constant and we get our anomalous scaling $\delta$ which is fitted with a straight line in a log-log representation. We choose to use 11 stripes for our analysis with DEA on all of our series generated by the model. Since we focus on the region $2 < \mu < 3$ for the analysis of our model we use the scaling relation 
\begin{equation} \label{scalingr}
\delta = \frac{1}{\mu-1}
\end{equation}
to relate the scaling $\delta$ given by the DEA to the value of $\mu$ we select for our model.
In our analysis we select an initial $\mu$ so that we know what scaling the DEA should recover when searching for crucial events. This is the same value of $\mu$ that we use for spacing the main shocks in equation (\ref{space}) in our model. For the case of overlapping sequences of aftershocks we find that mixing both kinds of non-stationarity in this way causes the analysis to give an inaccurate scaling regardless of the value of $\mu$. This dense mixing of non-stationarity creates a process that causes errors in recovering the proper scaling. While the DEA should give the scaling due to the presence of crucial events we find that a strong introduction of other forms of non-stationarity can create a process that hides the true scaling of the series. This discrepancy is shown in FIG. \ref{overlap}, which shows the scaling found when analyzing overlapping Omori sequences have unexpected values. These values diverge from those that should be found given our relation in equation (\ref{scalingr}). 

\begin{figure}
	\begin{center}
		\includegraphics[width=0.48\linewidth]{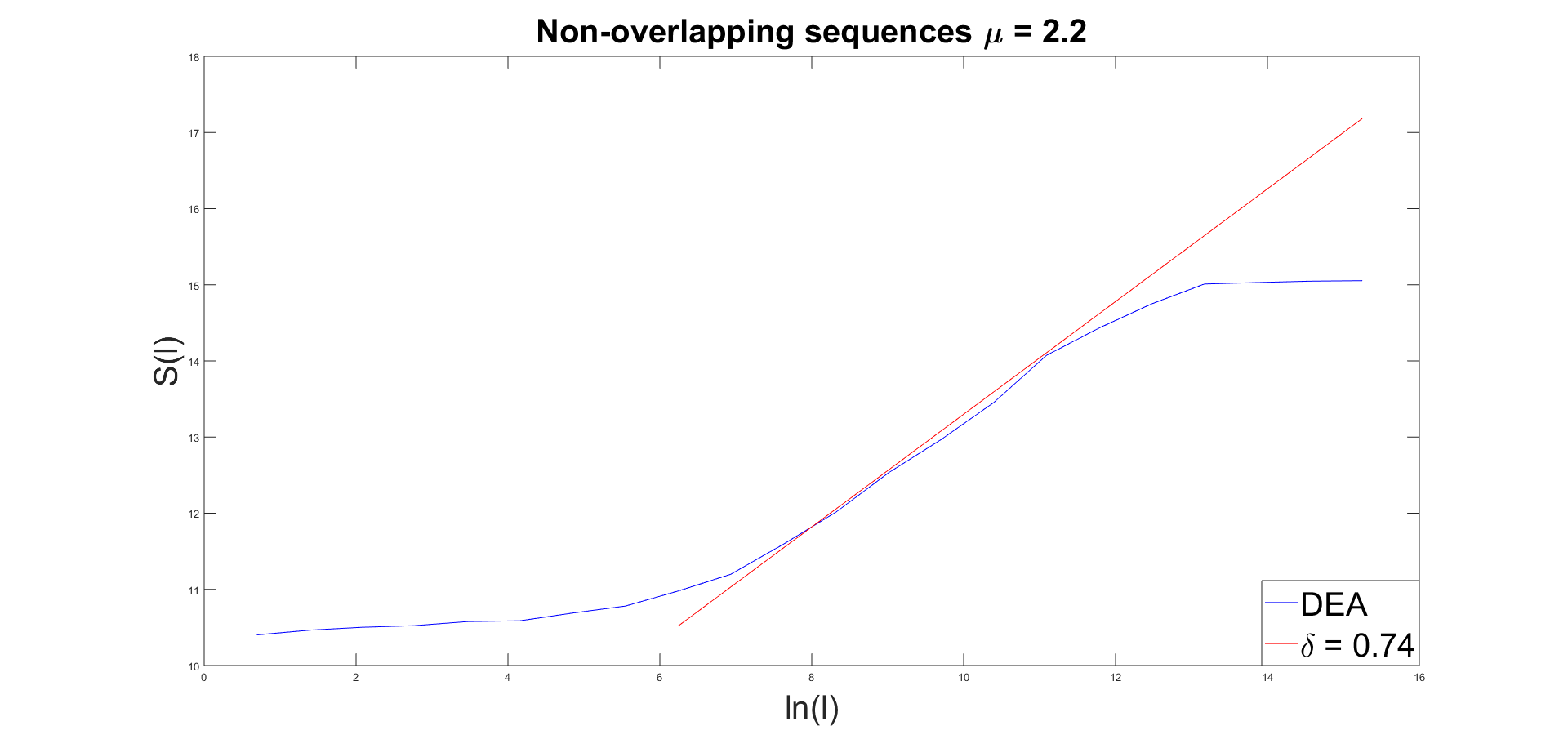}
		\includegraphics[width=0.48\linewidth]{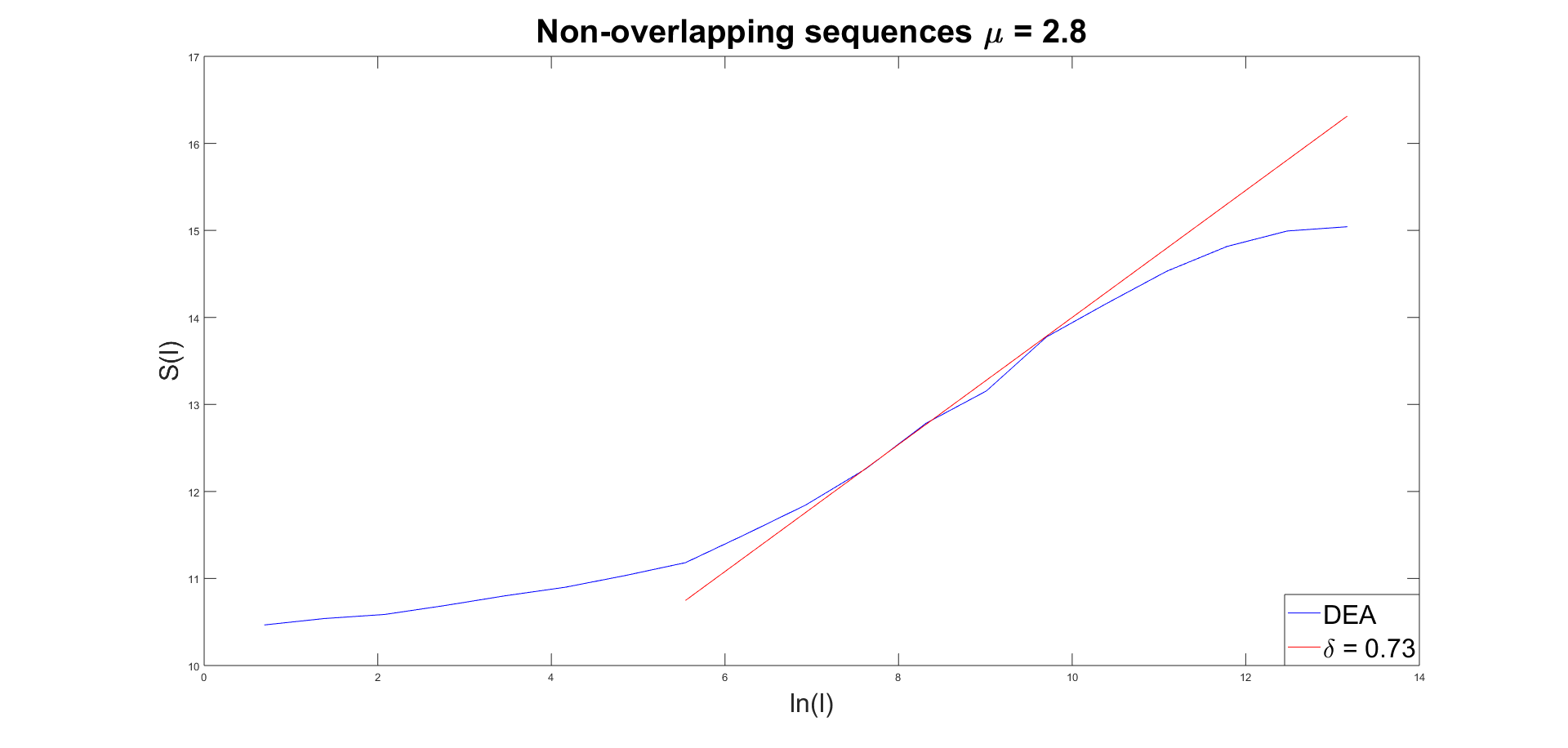}
		\includegraphics[width=0.48\linewidth]{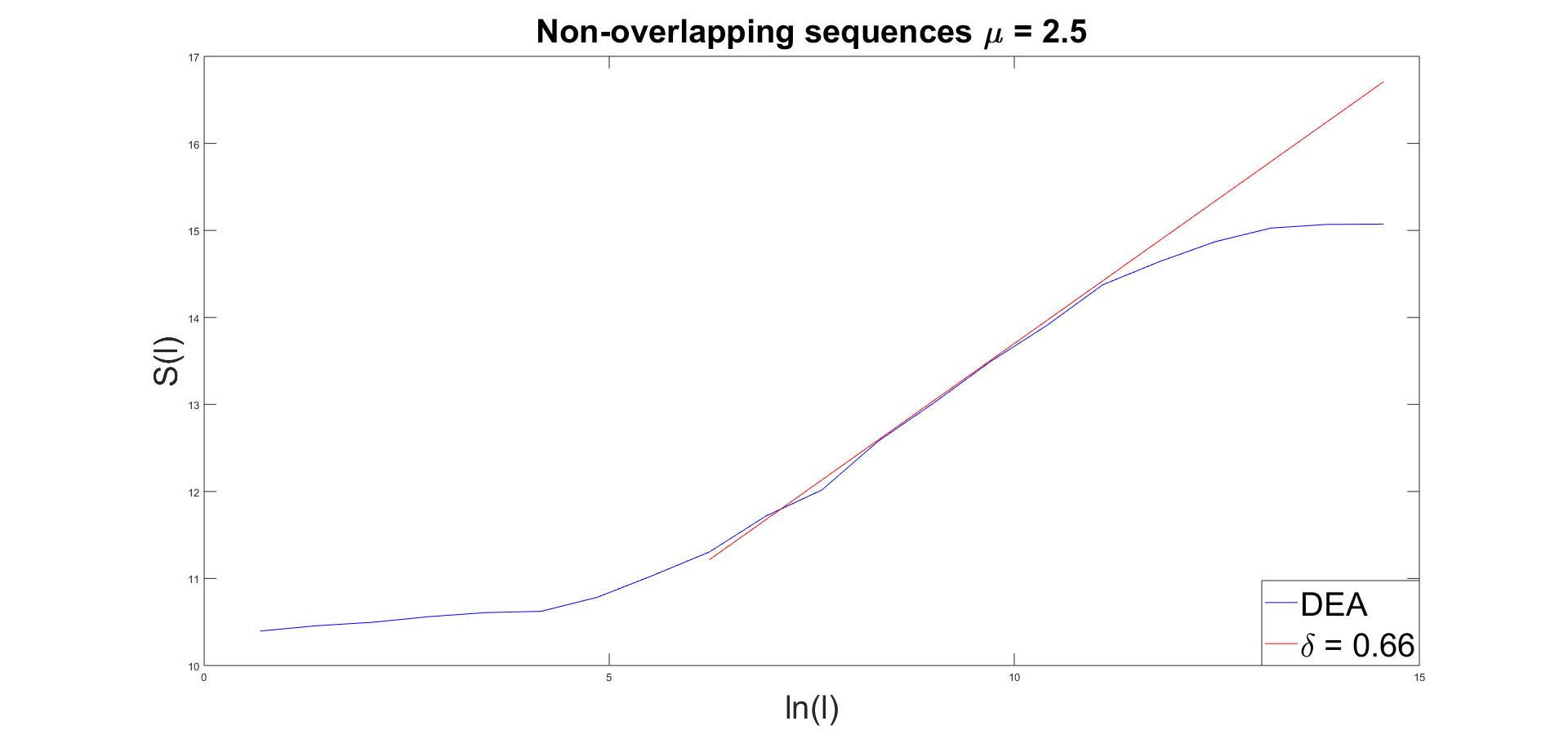}
		\includegraphics[width=0.48\linewidth]{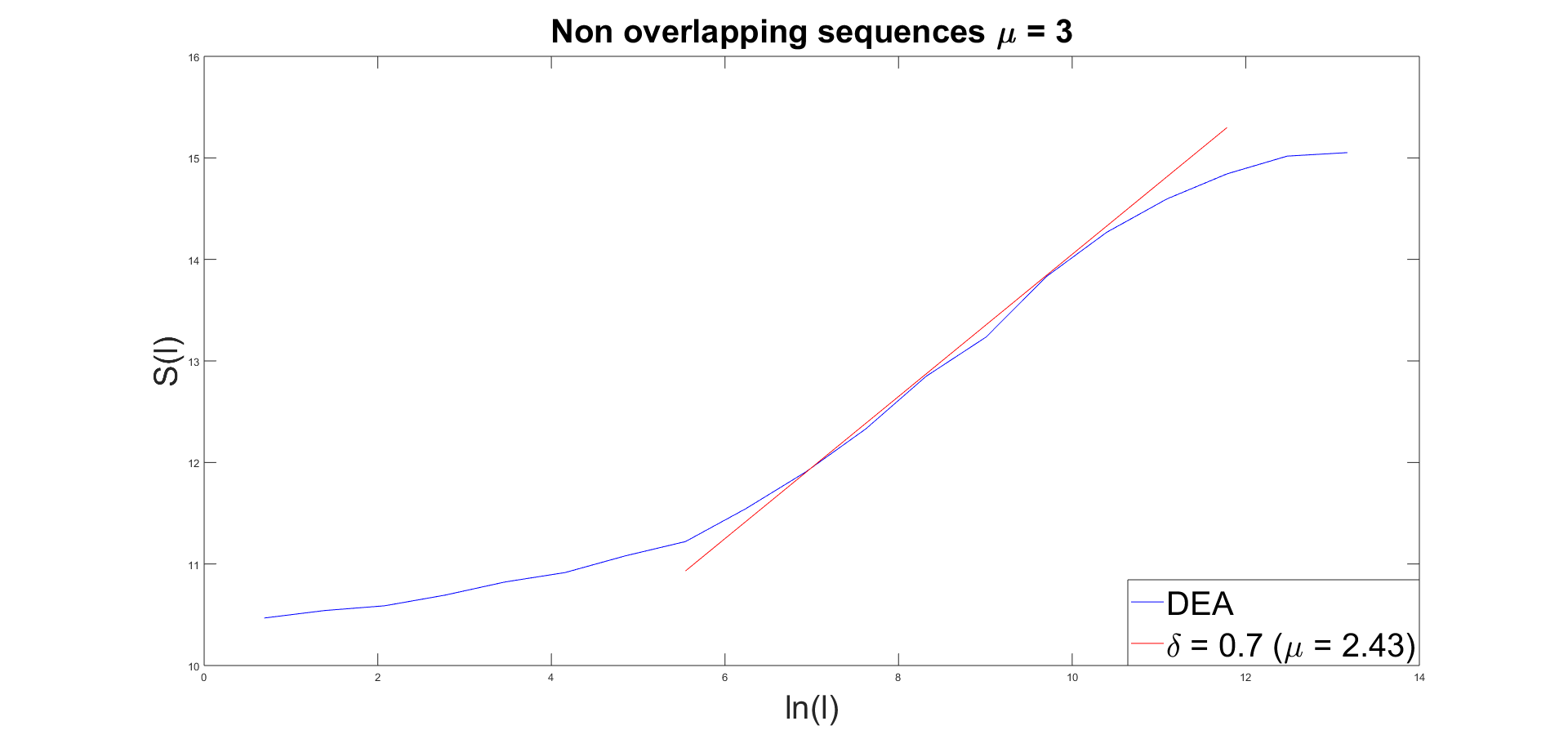}
		\caption{Analysis of various initial values of $\mu$ for non-overlapping sequences of aftershocks, making note of the perfect agreement for the value of $\mu = 2.5$. Titles include the initial value of $\mu$ and the legend provides the resultant scaling in a log-log plot.}
		\label{noverlap}
	\end{center}
\end{figure}

However, in the case of Omori sequences that are not allowed to overlap we find that we can recover the correct scaling for values of $\mu$ that are not close to the borders of our region of study. While the non-stationary elements are still both present, the intensity of the interaction is significantly weaker in this condition. This is shown in our Fig \ref{noverlap} for various values of $\mu$ where we see a perfect agreement at $\mu = 2.5$ and the divergences at values of $\mu$ close to the borders. 

This effect can be explained by noticing that both $\mu = 2$ and $\mu=3 $ are singularity conditions preventing the detection of crucial events in the presence of the perturbing action of Omori cascades, not only in the overlapping case but also in the non-overlapping one, if $\mu$ is not sufficiently far away from $\mu=2$ and $\mu= 3$.

In the case $\mu = 2$ we must recall the direction of \cite{raffaelli}. If we turn crucial events into a walk generating scaling $\delta$, the value of the scaling depends on the rule adopted to converting crucial events into a walk. In the case when the walker makes a step ahead of constant intensity when a crucial event occurs and it does not move if a crucial event does not occur,
then for $\mu < 2$

\begin{equation} \label{mu<2}
    \delta = \frac{\mu -1}{2}. 
\end{equation}

If $2 <\mu < 3$ we must use Eq. (\ref{scalingr}). We notice that the condition $\mu = 2$ is a kind of singularity that, in the presence of
the perturbation exerted by the occurrence of an Omori cascade, is very difficult to establish. In fact, in the long time limit we should find a distance from two consecutive crucial events leading to the correct value of $\mu$. The presence of events generated by an Omori cascade makes it impossible to find the correct value.

In the appendix \ref{iAppendix} we prove that the same additional term is present also for $\mu > 2$. This makes it difficult to establish the value at which the transition from the L\'{e}vy to the Gauss regime occurs. Actually, as proved in \cite{annunziato}, a sort of phase transition occurs at $\mu = 3$, forcing the width of the distribution density to generate a singularity. See Fig. 3 of Ref. \cite{annunziato}. Under the presence of an even weaker perturbation this singularity is replaced by a smooth transition. The Omori cascade exerts the same role as a weak perturbation making it impossible to detect crucial events, signalled by the slow tails of the L\'{e}vy distribution density.

This effect is in agreement with the results found in \cite{annunziato} at the border $\mu = 3$ and showing the discrepancy of theoretical prediction and numerical results in the region. We can also refer back to our discussion of section \ref{review} and see that our results match the observance of divergences at the borders of $\mu = 2$ and $\mu = 3$ as shown in equation (\ref{formauro}). The discrepancy at the borders can also be characterized by a phase change as well since at $\mu = 3$ there is a shift from Le\emph{v}y statistics to Gaussian as well as moving to a region characterized by perennial aging beyond $\mu = 2$. Ultimately we find that the strength of non-stationary interaction causes a shift in the detection of crucial events. This interaction strength along with the amplified effects of divergences at the borders show the optimal values of recovering correct information in areas defined by L\'{e}vy statistics. 

\section{concluding remarks} \label{last} 

\subsection{Theoretical problems with the Omori law}
The surprising result of this paper is illustrated in Fig.( {\ref{noverlap})} which shows that the adoption of DEA with stripes applied to the case of main shocks with $\mu = 2.5$ yields $\delta = 0.66$ which is the prediction of 
Eq. (\ref{scalingr}). In this case the detection of crucial events leads to an extremely accurate result. One may make the conjecture that the distance of $\mu = 2.5$ from the singularity of the state $\mu =2$ is large enough as to overcome the singularity effects of the phase transition at $\mu = 2$. However, we find that as we approach the value $\mu = 3$ this accuracy is lost and the value $\delta = 0.7$ that we find at $\mu = 3$ should correspond to a value of $\mu = 2.43$ according to Eq. (\ref{scalingr}). This surprising result is explained by the arguments of Section (\ref{3B}). In fact, in this section we find the quite surprising result of Eq. (\ref{equationofmauro}) which shows that at $\mu = 3$ the time necessary to reach the  stationary condition is infinitely large, because $1/t$ is at the border of integrability. These discrepancies can be caused by the assumption that an Omori cascade may last for an extended time period. While not long enough to reach a stationary condition, the long time correlation causes effects that result in the inability to properly detect the correct scaling of the sequence.

\subsection{Shedding light into a long lasting controversy}
To make easier for the readers to understand the importance of the conclusions of this paper we invite them to read Ref. \cite{comment} and \cite{reply}. In the model of \cite{comment} the authors found the same scaling value of $\delta =  0.9$ of \cite{mega} without making the assumption that the main shocks are crucial events. The  authors of \cite{reply} proved that this interesting result requires the action of an Omori cascade extremely extended in time, thereby involving again the action of a non-stationary process of second kind. We note that $\mu =2.1$ is a property shared by the neurophysiology of healthy brains \cite{2009}. On one side, we find it difficult to support the universality of non-stationary processes of the second kind to explain the observation of $\mu = 2.1$ in the dynamics of the underground, see \cite{mega} supplemented by the discussion of this paper on the interaction between non-stationary processes of first kind and non-stationary processes of the second kind. We also note that, in \cite{paper1}, if one does not focus on a non-stationary process of the second kind and instead on an interaction between the faults, one can achieve a scaling value closer to that found in real earthquake data. On the other side, having in mind also Ref. \cite{2009}, we are inclined to believe that $\mu =2.1$ is the manifestation of self-organized transitions to criticality, ranging from neurophysiology to fault interaction as originally proposed by Turcotte \cite{Turcotte}. A rigorous proof of this property that may be a remarkable contribution to current research work on the origin of "intelligence" would imply that the connection between main shocks and aftershocks is not yet understood, in agreement with the view of professional geophysicist \cite{poorlyunderstood}.


\appendix
\section{Rate event function}\label{iAppendix}
In this appendix we study the asymptotic behavior of the rate event function $R(t)$ in the particular case of the critical values of the power law $\mu=2,3$ 
corresponding to the appearance of the first and second finite moment. The Laplace transform of $R(t)$ is

\begin{equation}\label{pl1}
\hat{R}(s)=\frac{\hat{\psi}(s)}{1-\hat{\psi}(s)}  
\end{equation}
with $\hat{\psi}(s)$ Laplace transform of $\psi(t)$. We consider the convenient general expression in terms of $s$ power given by \cite{mauro}

\begin{equation}\label{trasf1}
\hat{\psi}(s)=-\Gamma(2-\mu)(s T)^{\mu-1}\left(\exp[s T]-E_{\mu-1}^{s T}\right)
\end{equation}
where

\begin{equation}\label{trasf2}
E_{\alpha}^{s T}=\sum_{n=0}^{\infty}\frac{(s T)^{n-\alpha}}{\Gamma(n+1-\alpha)}  
\end{equation}
Let us consider $\mu=2$ Eq. (\ref{trasf1}) is indeterminate. Using Taylor expansion for $\mu=2$ and taking into account that 

\begin{eqnarray}
\lim_{\alpha\to 0}\frac{\Gamma '(\alpha)}{\Gamma (\alpha)^2}=-1,\,\,\,\,\Gamma(2-\mu)\approx \frac {1}{2-\mu},
\end{eqnarray}
we have

\begin{eqnarray}\nonumber
&&E_{\mu-1}^{s T}\!\approx\! E_{1}^{s T}\!+\!\frac{\partial}{\partial \mu}  E_{\mu-1}^{s T}\mid_{\mu=2}\! (\mu-2)\!=\!\exp[sT]\!+\!\frac{\partial}{\partial \mu}  E_{\mu-1}^{s T}\mid_{\mu=2}\! (\mu-2)
\\\nonumber
&&=\exp[sT]+\left(\sum_{n=0}^{\infty}\frac{(s T)^{n-1}}{\Gamma(n)} \left[\frac {\Gamma'(n)}{\Gamma(n)}-\log(s T)\right]\right)(\mu-2)=
\\\label{trasf3}
&&\exp[sT]-\frac {1}{s}+\left(\sum_{n=0}^{\infty}\frac{(s T)^{n}}{\Gamma(n+1)} \left[\frac {\Gamma'(n+1)}{\Gamma(n+1)}-\log(s T)\right]\right)(\mu-2)
\end{eqnarray}
Taking the limit for $\mu\to 2$ in Eq. (\ref{trasf1}) and keeping the first two terms of the series we have

\begin{equation}\label{trasf4}
\hat{\psi}(s)\approx 1+s T \left[\log (sT)+\gamma\right], \,\,\,\, \gamma\equiv-\Gamma '(1)
\end{equation}
Consequently, in the limit $s\to 0$, we have

\begin{equation}\label{pl2}
\hat{R}(s)\approx \frac{\hat{\psi}(s)}{1-\hat{\psi}(s)}\approx -\frac{1}{s T(\log (sT)+\gamma )}\approx
\frac{1}{s T \log \left(\frac{1}{s T}\right)}
\end{equation}
Applying the tauberian theorem \cite{weiss} in general we have

\begin{eqnarray}\label{taub}
\hat{f}(s)= \frac{L\left(\frac{1}{s}\right)}{s^\alpha} \Rightarrow 
 f(t) = \frac{d}{d t} \left[\frac{t^\alpha L\left(t\right)}{\Gamma(\alpha+1)} \right]
\end{eqnarray}
with $L(t)$ a slow function. Using the result for $R(t)$ we have that

\begin{equation}\label{pl3}
R(t)\approx \frac{d}{d t} \left(\frac{t}{ \log t}\right)=\frac{1}{T \log \left(\frac{t}{T}\right)}-\frac{1}{T \log ^2\left(\frac{t}{T}\right)}\approx
\frac{1}{T \log \left(\frac{t}{T}\right)}
\end{equation}
Analogously, for $\mu=3$, we have (for $s\to 0$)

\begin{equation}\label{trasf5}
\hat{\psi}(s)\approx  1-sT -(sT)^2 ( \log (s T)+\gamma ) 
\end{equation}
and for the rate event

\begin{equation}\label{pl4}
\hat{R}(s)\approx 
\frac{1}{s T}- \log (sT)=\frac{1}{s T }+\log \left(\frac{1}{s T}\right)
\end{equation}
Applying the tauberian theorem

\begin{equation}\label{pl5}
R(t)\approx
\frac{1}{T}\left(1+\frac{T}{t}\right)
\end{equation}


\begin{references}

\bibitem{doublesornette} A. Sornette and D. Sornette, Self-Organized Criticality and Earthquakes,  EPL, {\bf 9}, 197 (1989). 

\bibitem{helmstetter} A. Helmstetter,  and D. Sornette, Diffusion of epicenters of earthquake aftershocks, Omori's law, and generalized continuous-time
random walk models, Phys. Rev. E, {\bf 66}, 061104 (2002). 

\bibitem{bak} P. Bak , K. Christensen , L. Danon , T. Scanlon T. 
Unified scaling law for earthquakes. Phys. Rev.
Lett. {\bf 88}, 178501 (2002). 

\bibitem{mega}  M. S. Mega, P. Allegrini , P. Grigolini, V.  Latora,
L. Palatella, A. Rapisarda, S. Vinciguerra S. Power-law time distribution of large
earthquakes. Phys. Rev. Lett. {\bf 90}, 188501 (2003). 

\bibitem{turcotte} R. Shcherbakov, G.  Yakovlev, D. L. Turcotte,  J.  B. Rundle, Model for the Distribution of Aftershock Interoccurrence Times, Phys. Rev. Lett., {\bf 95}, 218501 (2005). 


\bibitem{kagan}Y. Y.  Kagan, Earthquake size distribution:
power-law with exponent  $\beta = \frac{1}{2} ?$,
Tectonophysics {\bf 490}, 103 (2010). 


\bibitem{bacterial} E. Simsek, M. Kim,  Power-law tail in lag
time distribution underlies bacterial persistence.
Proc. Nat . Acad. Sci. USA {\bf 116}, 17635 (2019 ). 

\bibitem{solar}  D. E. Gary, C.U.  Keller,   Solar and space
weather radiophysics: current status and future
developments, vol. 314. Berlin, Germany:
Springer Science \& Business Media (2004). 

\bibitem{stok} X. Gabaix,  Power laws in economics
and finance. Annu. Rev. Econ. {\bf 1}, 255 (2009). 

\bibitem{tree} L. P. Bentley, J. C. Stegen, V.M. Savage ,  D.D. Smith,
E. I. von Allmen, J. S. Sperry, P.B. Reich , B.J. Enquist 
An empirical assessment of tree branching
networks and implications for plant allometric
scaling models. Ecol. Lett.{\bf 16}, 1069-1078 (2013). 


\bibitem{romanempire} P. L. Ramos, L. F. Costa F. Louzada and F. A. Rodrigues, Power laws in the Roman Empire: a survival analysis, R. Soc. Open Sci. 8:
210850 (2021). 

\bibitem{akimoto} T. Kataoka, T. Miyaguchi, T.  Akimoto, Detrended fluctuation analysis of earthquake data,  Phys. Rev. Res.{\bf 3}, 033081(2021). 


\bibitem{cox} D.R.Cox, Renewal theory, London: Methuen \& Co.Ltd.;1962.


\bibitem{feller}  W. Feller, An Introduction to Probability Theory and Its
Applications (Wiley \& Sons, New York, 1971).

\bibitem{bellazzini} P. Allegrini, J. Bellazzini, G. Bramanti, M. Ignaccolo, P. Grigolini, and J.Yang, Scaling breakdown:A signature of aging,
Phys. Rev. E 66, 015101(R) (2002).

\bibitem{barkai1} E. Barkai, Aging in Subdiffusion Generated by a Deterministic Dynamical System, Phys. Rev. Lett. 90, 104101 (2003).  


\bibitem{barkai2}  G. Margolin, E. Barkai, Nonergodicity of blinking nanocrystals and other Lévy-walk processes, Phys. Rev. Lett. 94, 080601 (2005).                     


\bibitem{barkai3}   E. Barkai, Residence time statistics for normal and fractional diffusion in a force field, J. Stat. Phys. 123, 883 (2006).      

            
\bibitem{barkai4}   G. Bel, E.Barkai, Weak ergodicity breaking with deterministic dynamics, Europhys. Lett. 74, 15 (2006). 


\bibitem{barkai5} G. Bel and E. Barkai, Occupation times and ergodicity breaking in biased continuous time random walks, J. Phys. Condens. Matter 17, S4287 (2005).          







\bibitem{annunziato}  M. Annunziato, P. Grigolini, Stochastic versus dynamic approach to L\'{e}vy statistic in the presence of an external perturbation,  Physics Letters A 269, 31 (2000). 

\bibitem{tenyearsearlier} P. Allegrini, Bellazzini, G. Bramanti,M. Ignaccolo,P. Grigolini, J. Yang, Scaling breakdown: A signature of aging, Phys. Rev. E, {\bf 66}, 015101 (R) (2002). 

\bibitem{hanngy} A. Rebenshtok, S. Denisov, P. Hanggi, E. Barkai, Infinite densities for L\'{e}vy walks, Phys. Rev. E, 90. 062135 (2014). 

\bibitem{metzler} P. Xu, R. Metzler, W. Wang, Infinite density and relaxation for L\'{e}vy walks in an external potential: Hermite polynomial approach, Phys. Rev. E 105, 044118 (2022). 

\bibitem{param} B. Choi, T. Itoi, T. Takada, Probabilistic aftershock occurrence model based on the 2011 Tohoku earthquake data, 
 World Conference on Earthquake Engineering  {\bf 15} (2012).

\bibitem{gutenomori} P.A. Reasenberg and L.M.Jones (1989). Earthquake hazard after a mainshock in California, Science, 243,
1173-1176
\bibitem{omori} F. Omori, On the aftershocks of earthquakes, J. Coll. Sci. Imp.
Univ. Tokyo 7, 111 (1894).

\bibitem{utsu} T. Utsu, Aftershocks and earthquake statistics II: Further inves-tigation of aftershocks and other earthquake sequence based
on a new classification of earthquake sequences, J. Fac. Sci.
Hokkaido Univ. Ser. VII 3, 197 (1970)

\bibitem{gutenberg} B. Gutenberg and C. F. Richter, Frequency of earthquakes in
California, Bull. Seismol. Soc. Am. 34, 185 (1944)

\bibitem{dea} G. Culbreth, B. J. West, P. Grigolini,  Entropic approach to the detection of crucial events. Entropy, 21(2), 178 (2019).


\bibitem{paper1} C. Muir, J. Cortez, P. Grigolini, Interacting faults in colifornia and hindu kush, Chaos, Solitons and Fractals, 139, 110070 (2020).

\bibitem{comment} A. Helmstetter, D. Sornette1,``Comment on ‘‘Power-Law Time Distribution of Large Earthquakes’’, Phys. Rev. Lett.,  92,129801 (2004). 

\bibitem{reply} M. S. Mega, P. Allegrini, P. Grigolini, V. Latora,
L. Palatella, A. Rapisarda, S. Vinciguerra, ``Mega et al Reply", Phys. Rev. Lett., {\bf 92},129802 (2004).

\bibitem{raffaelli} P. Grigolini, L. Palatella, G. Raffaelli, Asymmetric Anomalous Diffusion: an Efficient Way to Detect Memory in Time Series, Fractals {\bf 09},439 (2001).  

\bibitem{2009} P. Allegrini,D. Menicucci, R. Bedini, L. Fronzoni, A. Gemignani, P.  Grigolini,
B. J. West, P. Paradisi,  ``Spontaneous brain activity as a source of ideal$ 1/f$ noise, Phys. Rev. {\bf 86}, 061914 (2009).

\bibitem{Turcotte} D Turcotte, "Fractals, chaos, self‐organized criticality and tectonics." Terra Nova {\bf 4}, 4 (1992). 

\bibitem{poorlyunderstood} D. Zaccagnino, C. Doglioni, "Earth's gradients as the engine of plate tectonics and earthquakes", La Rivista del Nuovo Cimento, {\bf 45}, 801 (2022). 


\bibitem{mauro} B. J. West, M. Bologna,
P. Grigolini, "Physics of Fractal
Operators",
Springer-Verlag, New York, 2003.
ISBN 0-387-95554-2


\bibitem{weiss} G. H. Weiss, Aspects and Applications of the Random Walk (Random Materials and Processes), North Holland (1994).



\end{references}
\end{document}